\documentclass[prl,twocolumn,showpacs,preprintnumbers,amsmath,amssymb,superscriptaddress]{revtex4}

\usepackage{graphicx}% Include figure files
\usepackage{epsfig}
\usepackage{amsmath}
\usepackage{amssymb}
\usepackage{textcomp}
\usepackage{dcolumn}% Align table columns on decimal point
\usepackage{bm}% bold math
\usepackage{braket}
\usepackage[usenames,dvipsnames]{xcolor}
\usepackage[colorlinks=true,citecolor=MidnightBlue,linkcolor=MidnightBlue,urlcolor=MidnightBlue]{hyperref}

\PassOptionsToPackage{numbers,sort&compress}{natbib}

\makeatletter
\g@addto@macro\normalsize{%
  \setlength\abovedisplayskip{4pt}
  \setlength\belowdisplayskip{4pt}
  \setlength\abovedisplayshortskip{4pt}
  \setlength\belowdisplayshortskip{4pt}
}
\makeatother

\begin{document}

\newcommand*{\MAINZ}{QUANTUM, Institut f\"ur Physik, Universit\"at Mainz, Staudingerweg 7, 55128 Mainz, Germany}
\affiliation{\MAINZ}
\newcommand*{\ERLANGEN}{Institut f\"ur Optik, Information und Photonik, Universit\"at Erlangen-N\"urnberg, Staudtstr. 1, 91058 Erlangen, Germany}
\newcommand*{\SAOT}{Erlangen Graduate School in Advanced Optical Technologies (SAOT), Paul-Gordan-Str. 6, Universit\"at Erlangen-N\"urnberg,
	91052 Erlangen, Germany}
\homepage{http://www.quantenbit.de}

\title{Visibility of Young's interference fringes: Scattered light from small ion crystals}
\author{Sebastian Wolf}\email{wolfs@uni-mainz.de}\affiliation{\MAINZ}
\author{Julian Wechs}\affiliation{\ERLANGEN}
\author{Joachim von Zanthier}\affiliation{\ERLANGEN}\affiliation{\SAOT}
\author{Ferdinand Schmidt-Kaler}\affiliation{\MAINZ}

\date{\today}% It is always \today, today,

\begin{abstract}
We observe interference in the light scattered from trapped $^{40}$Ca$^+$ ion crystals. By varying the intensity of the excitation laser, we study the influence of elastic and inelastic scattering on the visibility of the fringe pattern and discriminate its effect from that of the ion temperature and wave-packet localization. In this way we determine the complex degree of coherence and the mutual coherence of light fields produced by individual atoms. We obtain interference fringes from crystals consisting of two, three and four ions in a harmonic trap. Control of the trapping potential allows for the adjustment of the interatomic distances and thus the formation of linear arrays of atoms serving as a regular grating of microscopic scatterers.

\end{abstract}

%  37.10.Ty   Ion trapping
%  37.10.Vz   Mechanical effects of light on atoms, molecules, and ions
%  42.50.Ct   Quantum description of interaction of light and matter; related experiments

\maketitle

The seminal double slit experiment by Young \cite{shamos1959great} is one of the most prominent experiments in physics. Originally, it formed the basis for understanding that light is a wave giving rise to phenomena like interference and diffraction, whereas in its modern interpretation it displays in a compact form the notion of wave-particle duality \cite{Dirac1989}.

The original Young experiment employed transversally coherent light using a small aperture placed in front of the light source (in fact the sun~\cite{shamos1959great}). This results in electromagnetic waves at the two slits oscillating in phase and a visibility of the fringe pattern $\sim 100\%$. The use of laser-driven atoms as ``slits'' enables the formation of more complex light fields, ranging from fully coherent to partially coherent and even fully incoherent fields. This transition arises from the fundamental process of photon scattering by the atoms. In the quantum theory of light \cite{LOUD2000,Cohen2004,Scully1997} the scattering event involves the destruction of an incoming photon and the creation of an outgoing photon. For low intensities the elastic process dominates such that the outgoing photon has the same frequency and a fixed phase relationship with the incoming one \cite{Walther1987,Eschner2001}. Interferences in this regime have been observed in a seminal experiment by Wineland and coworkers involving two mercury atoms trapped in an ion trap and only weakly excited by a near-resonant laser \cite{EICH93} (see also \cite{itano1998complementarity,Ficek2005,Brewer1996,Walls1997,Skornia2001,Beige2001}). 

However, when increasing the intensity of the laser, the atomic emitters undergo internal dynamics which may alter the emitted photon frequency and phase. Such inelastic scattering processes lead to a reduced mutual coherence of the light fields, i.e., the emission of partially coherent light, resulting in a decrease of the visibility of the interference fringes. In the case of a very intense driving laser, the atoms emit fully incoherent fluorescence light \cite{Mollow1969}; in this case the visibility of the fringe pattern disappears. 

Aside from the internal dynamics, the driving laser affects additionally the external degrees of freedom of the ions as the laser is used likewise for laser cooling of the particles. The ion temperature plays an important role for the fringe visibility as it determines the localization of the scatterers, i.e., of the ``slits''. Since an increased laser intensity alters both the ratio of elastic to inelastic scattering as well as the localization of the atoms, the influence of inelastic scattering on the mutual coherence of the scattered light has not been observed experimentally.

In this letter we study the visibility of Young interference fringes produced by individual atoms employing a gated detection method to clearly separate the effect of inelastic scattering from that of reduced atom localization. A theoretical model to explain the measured fringe patterns is developed taking into account the multi-level structure of the atoms and the presence of a repumping laser.  Experimentally, we investigate ion crystals with up to four ions in a harmonic trap potential or in specially shaped trapping fields that allow for the adjustment of the interatomic distances. In this way we are able to form linear arrays of ions serving as a regular grating of atomic scatterers. 

\begin{figure}[htp]\begin{center}
\includegraphics[width=0.45\textwidth]{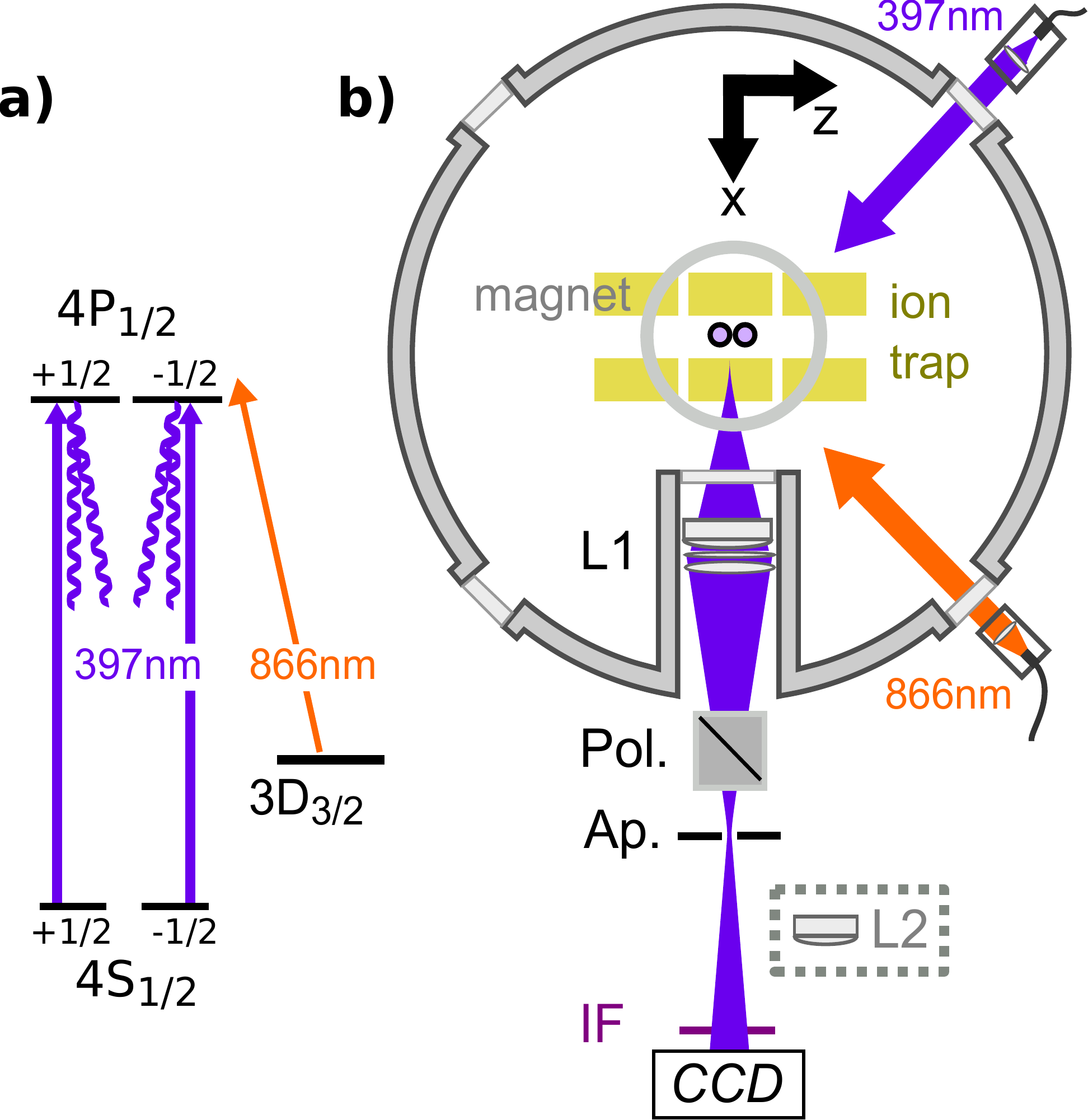}
\caption{(a) Level scheme and relevant transitions of the $^{40}$Ca$^+$ ion including the metastable 3$^2$D$_{3/2}$ state. (b) Sketch of experimental setup: Ions are held in a segmented micro trap (yellow), forming linear crystals along the z-axis, and are illuminated by laser light near 397~nm and 866~nm (for details see text). }
\label{Fig1_aufbau}
\end{center}\end{figure}

For the experiments we employ $^{40}$Ca$^+$ ions trapped in a segmented Paul trap \cite{Jacob2014}. With trap frequencies $\omega_{r_1,r_2, z}/(2\pi$)=(1.853, 2.620, 0.977)~MHz the ions form linear crystals which align along the weakest trap axis $\textbf{e}_z$. The electric dipole transition 4$^2$S$_{1/2} \rightarrow$ 4$^2$P$_{1/2}$ of $^{40}$Ca$^+$  near 397~nm is used for Doppler cooling and light scattering. The 4$^2$P$_{1/2}$ state decays with a probability of 7\% to the metastable 3$^2$D$_{3/2}$ level \cite{Hettrich2015}, therefore we use a laser near 866~nm for repumping to maintain continuous Doppler cooling (see Fig.~\ref{Fig1_aufbau}a). The radial modes $\omega_{r_1,r_2}$  are aligned along the $\textbf{e}_{\pm x + y}$ direction, respectively, whereas the cooling and repumping laser illuminate the ion crystals along the (x,y,z)=($\pm$1,0,-1)$/\sqrt{2}$ direction, respectively, so that the $\textbf{k}$-vectors of the laser beams have a projection on all vibrational axes of the ion crystal (see Fig.~\ref{Fig1_aufbau}b). 

A magnetic field of $\sim 0.24$~mT oriented along $\textbf{e}_y$, generated by a permanent magnet ring placed on top of the vacuum chamber, determines the quantization axis. The laser beam near 397~nm, having a waist of about $600$~$\mu$m at the ions' positions, is linearly polarized along this axis and thus excites the $\Delta m =0$ transitions (see Fig.~\ref{Fig1_aufbau}a). The light scattered  by the ions is collected by a f/1.6 objective L1 (focal length 67~mm) at a working distance of 48.5~mm and focused at a distance of about 770~mm, after being sent through a polarization beam splitter (Pol.) oriented along $\textbf{e}_y$, i.e., the same axis as the cooling laser (see Fig.~\ref{Fig1_aufbau}b). An aperture (Ap.) (diameter $\sim$~400 $\mu$m) is placed at the back focal plane of the objective suppressing unwanted stray light in combination with an infrared filter (IF,  center wavelenght $\lambda = 394 \pm 10$~nm). The scattered light is finally recorded by  a CCD camera positioned $\sim$ 100~mm behind the back focal plane of the objective to observe the light in the far field, i.e., the Fourier plane of the ions. We use either an  electron multiplier gain intensifier enhanced CCD camera (EMCCD, Andor iXon 860) or alternatively an intensified CCD camera (ICCD, Andor iStar 334T) with $128 \times 128$ pixels (pixel size 24.5~$\mu$m) and $1024 \times 1024$ pixels (pixel size 13~$\mu$m), respectively. A lens L2 (focal length f=25~mm), optionally placed in the scattered light beam behind the aperture, focuses the back focal plane onto the CCD allowing one to image and observe the ions individually, e.g. to check for the number of ions, to determine the magnification of the optical system or to adjust the axial potential. 

\begin{figure}[htp]\begin{center}
\includegraphics[width=0.45\textwidth]{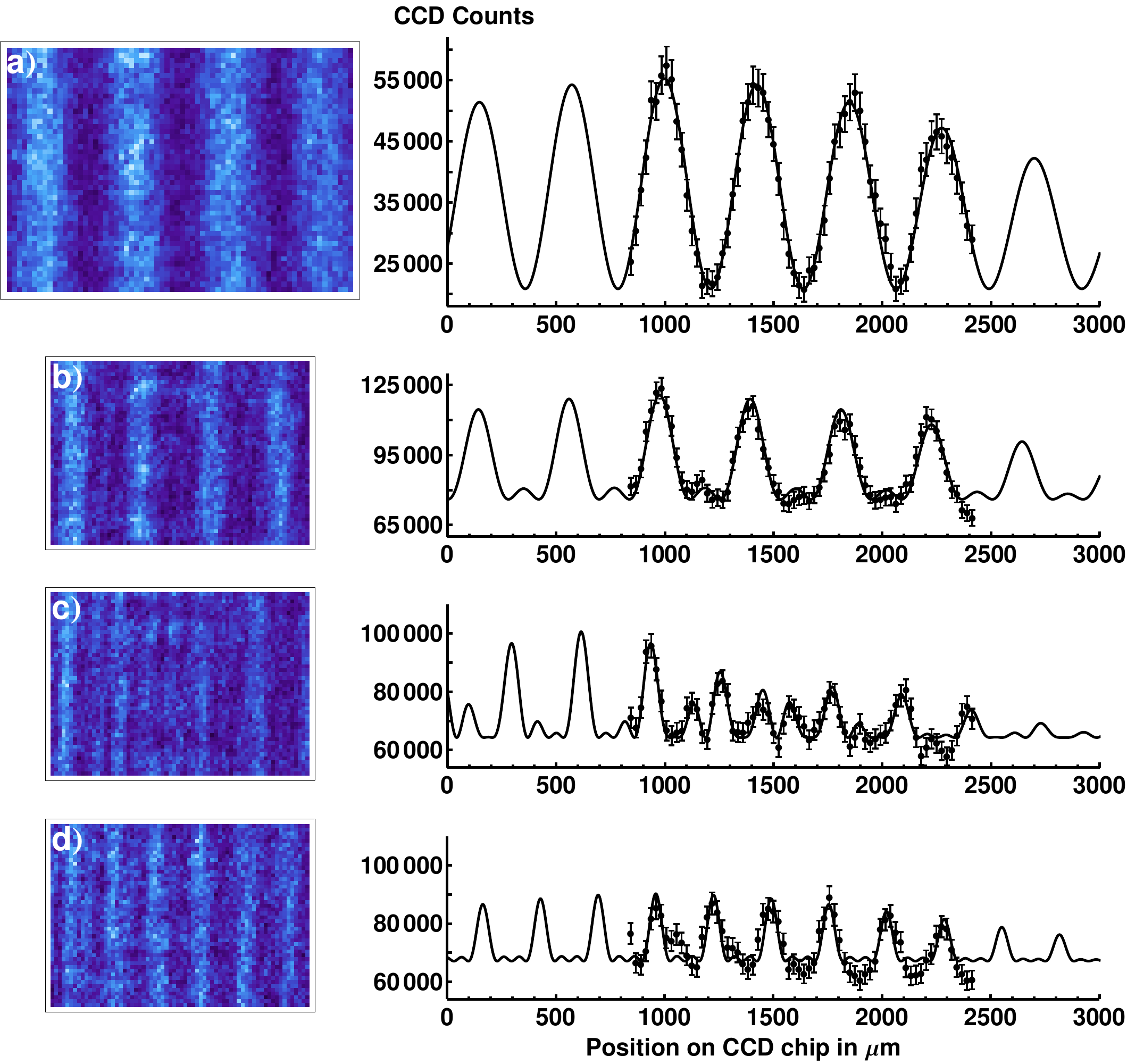}
\caption{Images of the EMCCD camera (left) and interference fringe patterns (right) for a) two b) three c) four ions in a harmonic trap potential. In d) data is presented for a crystal with four equidistant ions. The EMCCD images have been rotated, distortion-corrected and the background has been subtracted (for details see text). Note that the data of the EMCCD camera include the internal avalanche gain and is integrated over an exposure time of 60~s. The fringe patterns are obtained from the corrected EMCCD images by integration over the vertical axis.  Errors on each data point correspond to photon shot noise, dark noise and read-out noise. From a fit of the experimental curves we obtain a visibility $\mathcal{V}$ of the fringe patterns of $45.2 (6) \%$, $22.7 (6) \%$, $22 (1) \%$ and $15 (1) \%$ for the two-, three-, four-ion crystal, and the equidistant four-ion array, respectively, where the errors represent the root mean square deviation of each fit.}

\label{Fig2_fringes}
\end{center}\end{figure}

The results of the interference measurements for two, three and four ions are shown in Fig.~\ref{Fig2_fringes}. The inner parts of the CCD images ($68 \times 48$ pixels) are rotated and corrected for field distortions measured independently by observing the distance of a two-ion crystal at different positions within the field of view of the CCD. Remaining stray light and background are subtracted from the CCD images, determined by shutting off the repumping laser. The fringe patterns at the right hand side of Fig.~\ref{Fig2_fringes} are obtained from the CCD images by integration over the vertical axis; the error bars of $\sim 5 \%$ are deduced from photon shot noise. The fits to the interference patterns are derived from the source distribution via Fourier transformation, taking into account the resolution of the imaging device. From the fit parameters we determine the distance $d$ between the ions, the width $w$ of the point spread function (PSF), and the visibility $\mathcal{V}$ of the interference fringes. From Fig.~\ref{Fig2_fringes}a, we obtain a distance $d=6.4~\mu$m and a width of the PSF $w = 3.6~\mu$m for the two-ion crystal. Note, that the calculated magnification of the optical system - derived from the image of the back focal plane of L1 on the CCD by use of L2 - depends on the exact x-position of L2 which can be positioned with an accuracy of $\sim$~2 mm. In view of this uncertainty we see good agreement of the determined value $d$ with the independently deduced $d_{theo.} = 5.8~\mu$m, based on (i) a spectroscopic determination of the COM-mode frequency of the crystal and (ii) the calculation according to \cite{james1998quantum}. 

Key for the further studies is the gated cooling probe detection (GCPD) of the scattered photons made possible by our intensifier enhanced CCD camera. 
The GCPD scheme works as follows (see Fig. 3): The ion crystals are initialized during 175~$\mu$s via Doppler cooling under optimum conditions for the saturation $s_{397}$ and $s_{866}$ of the cooling and repumping lasers at 397~nm and 866~nm, i.e., well below the respective saturation intensities, and with a cooling and repumping laser detuning of $\Delta_{397}= -10$~MHz and  $\Delta_{866} = + 60$~MHz, respectively. We choose the laser detuning for the laser at 866~nm to the blue side of the resonance in order to avoid complications from dark resonances. Thereafter the saturation of the cooling laser $s_{397}$ is switched to a different value using an acousto-optical modulator. After a delay  of 5~$\mu$s to allow for proper switching of the laser, the CCD is gated for 10~$\mu$s to observe the scattered light at 397~nm.  As the motional states of the ion crystals evolve over much longer time scales (see Fig. 4), they are unable to adapt to the modified cooling laser saturation  within this detection time. In  this way the mutual coherence of the scattered light fields is solely determined by the internal degrees of freedom of the ions.  
We can thus investigate the visibility of the interference pattern as a function of the laser saturation  without being affected by the ion temperature. 

In the paraxial approximation and for scalar fields, i.e., for identical polarization of excitation and detection, the intensity produced by a two-ion crystal at the CCD is~\cite{Mandel}
\begin{equation}
\label{equation_1}
\begin{split}
I ( \textbf{r}, t) = &
 I_1 ( \textbf{r}, t) + I_2 ( \textbf{r}, t) + \\
&  2 \sqrt{ I_1 ( \textbf{r}, t) } \sqrt{ I_2 ( \textbf{r}, t) } \operatorname{Re}\left\{ \gamma ( \textbf{r}_1, \textbf{r}_2, \tau) \right\},
\end{split}
\end{equation}
where $I_1 ( \textbf{r}, t) $ ($I_2 ( \textbf{r}, t) $) is the intensity at $\textbf{r}$ if ion 2 (ion 1) is absent, $\operatorname{Re} \left\{  . \right\}$ denotes the real part and  $\varphi = k \, c \, \tau = k \, (|\textbf{r} - \textbf{r}_1| - |\textbf{r} - \textbf{r}_2|) $ is the relative phase accumulated by the fields  at $\textbf{r}$. 
In Eq.~(\ref{equation_1}), $\gamma ( \textbf{r}_1, \textbf{r}_2, \tau) = \langle E_1(\textbf{r}_1,t-\tau) E_2^*(\textbf{r}_2,t) \rangle / \sqrt{\langle |E_1(\textbf{r}_1)|^2 \rangle \langle |E_2(\textbf{r}_2)|^2 \rangle}$ corresponds to the complex degree of coherence which describes the mutual coherence of the two light fields $ E_1(\textbf{r}_1,t)$ and $ E_2(\textbf{r}_2,t)$, generated by ion 1 at $\textbf{r}_1$  and ion 2 at $\textbf{r}_2$, respectively.
We assume identical excitation strength and thus equal intensities $I_1 (\textbf{r}, t) = I_2 ( \textbf{r}, t) \equiv I_0$ of the two ions. The visibility of the interference fringes is then equal to the modulus of the complex degree of coherence and the fringe modulation determined by the phase $\varphi$.

In a three-level model and with the ions at fixed positions, the intensity distribution on the CCD is (see Supplemental Material)
\begin{equation}
\label{equation_3}
I(\textbf{r})= 2 \, I_0 \, ( 1 + |\rho_{sp}|^2/\rho_{pp} \cos{\varphi}) \; ,
\end{equation}
where $\rho_{sp}$ denotes the single atom coherence between states s = S$_{1/2}$ and p = P$_{1/2}$, and $\rho_{pp}$ is the population of the excited state decaying either to s or level d = D$_{3/2}$. According to Eq.~(\ref{equation_3}) the visibility of the interference pattern is given by
\begin{equation}
\label{equation_2}
\mathcal{V} = |\gamma| =  | \rho_{sp} |^2/\rho_{pp} \; .
\end{equation}
A reduction of $\mathcal{V}$ is thus predicted for growing $\rho_{pp}$ and reduced $\rho_{sp}$. If we model the ions as two-level atoms (for which  Eqs.~(\ref{equation_3}) and (\ref{equation_2}) equally hold) this occurs for increased laser saturation $s_{397}$. However, the two-level model does not take into account the modification of $\rho_{sp}$ and $\rho_{pp}$ due to the  additional decay channel to d. In this case $\rho_{sp}$ and $\rho_{pp}$, and thus Eq.~(\ref{equation_2}), become more involved functions of the laser parameters.

\begin{figure}[htp]\begin{center}
\includegraphics[width=0.47\textwidth]{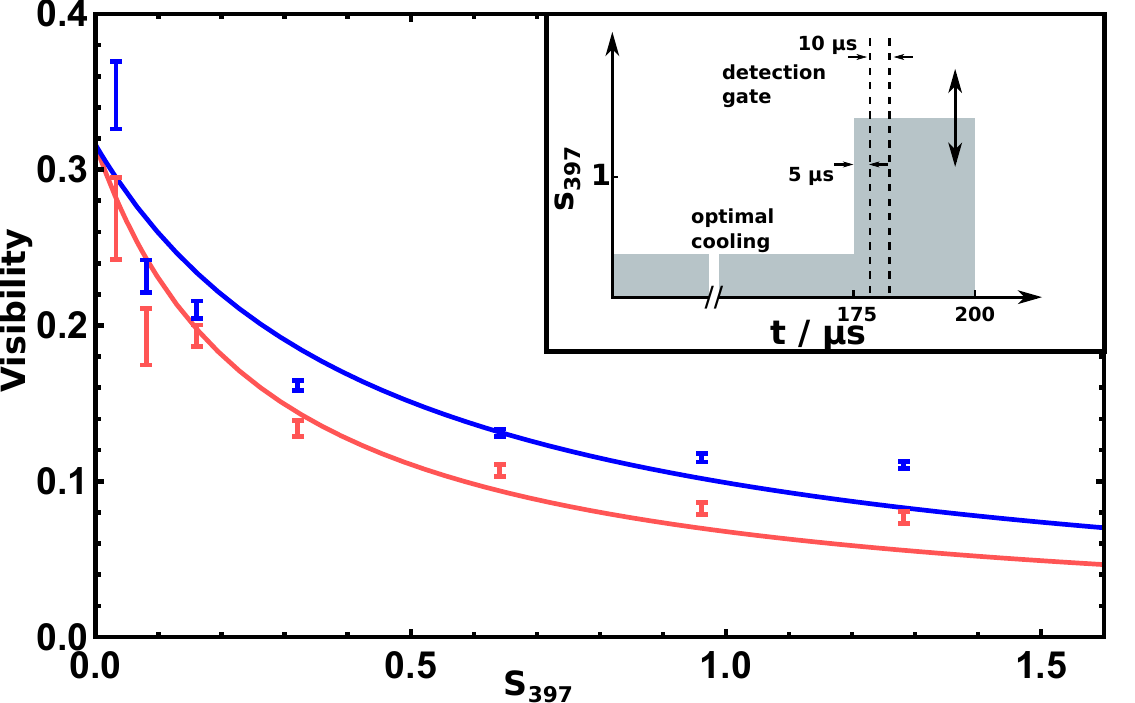}
\caption{Interference fringe visibility at the crossover of elastic to inelastic scattering for a two-ion crystal as a function of laser saturation $s_{397}$. The repumping laser saturation corresponds to s$_{866}$= 0.032 (red dots) and s$_{866}$= 0.15 (blue squares). As in Fig.~\ref{Fig2_fringes} vertical error bars represent the root mean square deviation of each fit. 
Note that the data has been obtained after the setup was modified to allow for the GCPD technique. Thus several experimental parameters are not exactly equal to Fig.~\ref{Fig2_fringes}, including a possible misalignment of the quantization axis.
The conversion from the measured laser powers into saturation involves the knowledge of the laser waists $\varnothing_{397} = 600 (\pm 300) \, \mu$m, $\varnothing_{866} = 300 (\pm 150) \, \mu$m and laser detunings $\Delta_{397}= -10 $~MHz, $\Delta_{866}= +60 (\pm 10)$~MHz.  All listed uncertainties lead to a systematic uncertainty of s$_{397}$ of about 100\%.
}
\label{Fig3_visibility}
\end{center}\end{figure}

The measured $\mathcal{V}$ produced by two-ion crystals as a function of $s_{397}$ is shown in Fig.~\ref{Fig3_visibility}. A reduction of $\mathcal{V}$, corresponding to the emission of {\it partially coherent light}, is observed when increasing $s_{397}$, which agrees well with the two-level model. When the saturation  of the repumping laser is increased by a factor of $\sim 4$ we observe, however, an increased visibility. This behavior is well described by the three-level model fit curves in Fig.~\ref{Fig3_visibility} (see Supplemental Material).

The visibilities displayed in Fig.~\ref{Fig3_visibility} are limited by a constant prefactor of $\sim$ 0.3.
Assuming this factor is only due to the motional excitation of the ion crystal results in a mean wavepacket size (r.m.s. of breathing and rocking modes) of 96(5) nm \cite{itano1998complementarity}. This is, however, about a factor 2.3 larger than that expected for the Doppler cooling limit, calculated for the given trap frequencies and unsaturated cooling; we measured a mean wavepacket size of 42(11)~nm using sideband spectroscopy \cite{Haeffner2008}.
We suspect therefore that the prefactor is also affected by misalignment of the quantization axis.

\begin{figure}[htp]\begin{center}
\includegraphics[width=0.47\textwidth]{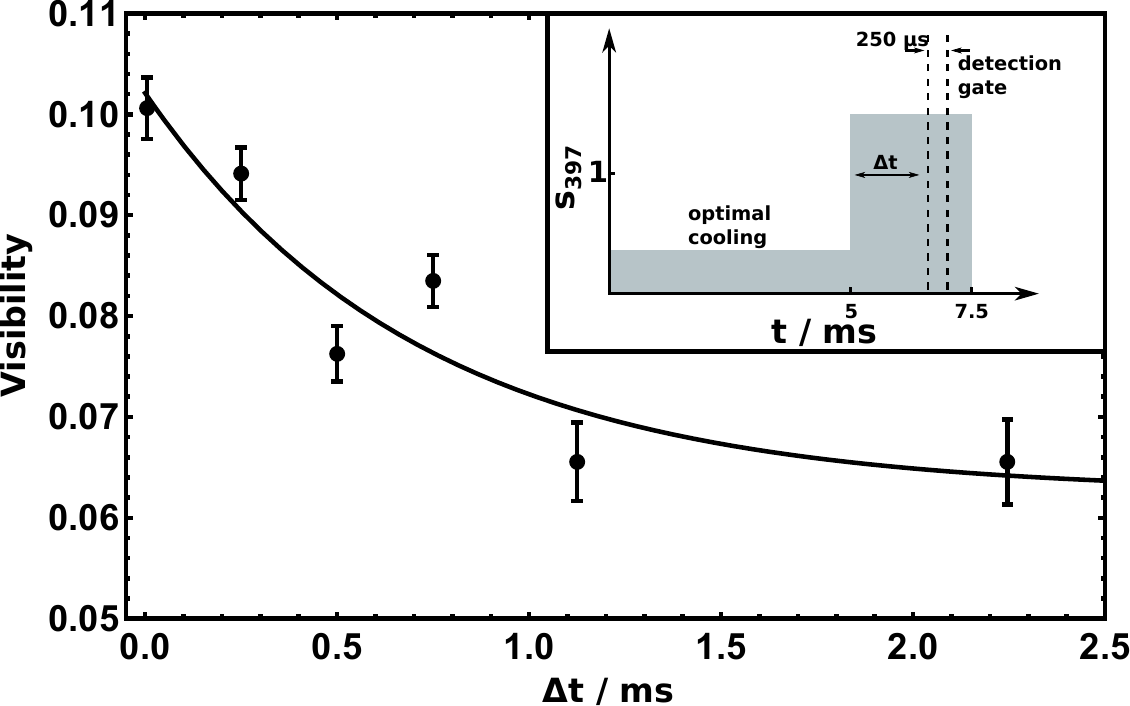}
\caption{Dynamical change of the interference fringe visibility $\mathcal{V}$ when heating up the two-ion crystal (for details and explanation of the inset see text). 
}
\label{Fig4_temp}
\end{center}\end{figure}

The GCPD scheme can also be employed to investigate the modification of the  fringe visibility due to vibrational excitations in the ion crystal.
Again, we initialize the crystal by Doppler cooling under optimum conditions (s$_{397}\sim0.25$, s$_{866}\sim0.16$). The laser saturation $s_{397}$ is then rapidly increased by a factor of $\sim 5$ while keeping the detuning unchanged. Here the CCD is gated to observe the scattered photons in a time interval of 250~$\mu$s while we shift the beginning of this time interval from $\Delta t = 0$~ms to 2.5~ms (see inset of Fig.~\ref{Fig4_temp}). As the crystal is exposed to a higher saturation, the Doppler cooling limit and the mean phonon number in the breathing and rocking modes increases \cite{EICH93}.
The visibility of the fringe pattern is proportional to the Debye Waller factor $\exp\{-\frac{1}{2}\langle (\textbf{k}_{\textrm{eff}} \cdot (\textbf{u}_1 - \textbf{u}_2))^2 \rangle \} $, where $\textbf{u}_i$ denotes the fluctuation about the equilibrium positions of ion i=1,2, $\textbf{k}_{\textrm{eff}}$ is the k-vector difference of the absorbed and emitted photons and $\langle \; \rangle$ denotes the average over the thermal distributions \cite{itano1998complementarity}. 

In the experiment the decrease of the fringe visibility as a function of $\Delta t$ is clearly visible (see  Fig.~\ref{Fig4_temp}), following an exponential decay with a time constant $\tau= 0.7(4)~$ms. The long time constant confirms our assumption that the time evolution of internal and external degrees of freedom of the ions can be separated by use of the GCPD approach. We have obtained similar data for the increase of  $\mathcal{V}$ when an initially higher crystal temperature is reduced by Doppler cooling.

Modern trap technology \cite{Brown2012,Home2009}, where the DC trap potential is shaped by multiple control segments, allows one to modify the trap potential along $\textbf{e}_z$ and thus the inter-ion distances. This becomes particularly relevant for crystals with $\ge 4$ ions. If a crystal with four ions is kept in a harmonic trap the equilibrium positions of the ions are non-equidistant~\cite{james1998quantum}, e.g., for trap frequencies $\omega_{r_1,r_2, z}/(2\pi$)=(1.978, 2.180, 0.429)~MHz the distance between the innermost ions is 7.2~$\mu$m and between the outer and the inner ions 7.6~$\mu$m,  respectively. This results in an interference fringe signal with two spatial frequencies (see  Fig.~\ref{Fig2_fringes}c). By adjusting the trap control electrode voltages we are able to generate a non-harmonic potential \cite{Ruster2014} such that a regular crystal with equal ion separation of 9.1~$\mu$m is obtained (see Fig.~\ref{Fig2_fringes}d). The corresponding fringe pattern matches the intensity distribution of a coherently illuminated 4-slit grating.

In conclusion, we studied the mutual coherence of light fields emitted by individual atoms at the crossover from elastic to inelastic scattering. We implemented a detection scheme allowing to observe the degree of mutual coherence as a function of the saturation of the observed S$_{1/2} \rightarrow$ P$_{1/2}$ transition at fixed ion crystal temperatures. The decrease of the visibility of the interference patterns due to motional effects of the ions was investigated separately. The method could pave the way towards temperature measurements of ion crystals at low trap frequencies where standard sideband methods, highly successful in tightly confining potentials \cite{Haeffner2008}, become increasingly hard. We also see applications when the trap potential is adiabatically lowered \cite{poulsen2012efficient}, e.g., when ions are loaded into optical potentials \cite{Schneider2010,Schmiegelow2015,Raey2014}.
The experiment also provides opportunities to investigate multi-ion entanglement 
\cite{Thiel2007,Bastin2009,Monroe2007,Rempe2012,Weinfurter2012,Hanson2013} or measurements of photon-photon correlations 
and their back action on the ion crystals 
\cite{Oppel2012,Genovese2014,Oppel2014}.     

\begin{acknowledgments}
We gratefully acknowledge the support of LOT\hbox{-}QuantumDesign for lending the ICCD. We thank S.T. Dawkins for carefully reading the text. FSK and SW acknowledge the financial support of the Cluster of excellence PRISMA at the Johannes-Gutenberg Universit\"at Mainz and the DFG within the project BESCOOL. JvZ gratefully acknowledges funding by the Erlangen Graduate School in Advanced Optical Technologies (SAOT) by the German Research Foundation (DFG) in the framework of the German excellence initiative.
\end{acknowledgments}

\bibliographystyle{apsrev4-1}
\bibliography{lit}

\newpage
\onecolumngrid

\section{Supplemental Material: Visibility of the interference pattern}

\subsection{Optical Bloch equations}
The interaction of an atom with laser light can be described by the optical Bloch equations \cite{cohen}.
For a two-level system with a ground state $\ket{s}$ and an excited state $\ket{p}$ and an exciting laser field with Rabi frequency $\Omega_{397}$ and detuning $\Delta_{397}$, they read \\
\begin{equation}
\begin{aligned}
&\dot{\tilde{\rho}}_{\text{ps}} = -(\frac{\Gamma_{\text{ps}}}{2} - i \Delta_{397}) \tilde{\rho}_{\text{ps}}-\frac{i}{2} \Omega_{397} (\rho_{\text{pp}}-\rho_{\text{ss}}) \\
&\dot{\rho}_{\text{pp}} = - \Gamma_{\text{ps}} \rho_{\text{pp}} + i \frac{\Omega_{397}}{2}(\tilde{\rho}_{\text{sp}}-\tilde{\rho}_{\text{ps}})\\
&\tilde{\rho}_{\text{sp}}=\tilde{\rho}_{\text{ps}}^{*}\\
&\rho_{\text{ss}}+\rho_{\text{pp}} = 1
\end{aligned} 
\end{equation}
where $\Gamma_{\text{ps}}$ is the decay rate of the s-p-transition and $\tilde{\rho}_{\text{ps}} = e^{i \Delta_{397} t}\rho_{\text{ps}}$. \\ 
In the three level system under investigation, there is an additional metastable state $\ket{d}$ and a laser field with Rabi frequency $\Omega_{866}$ and detuning $\Delta_{866}$ driving the $\ket{d}$-$\ket{p}$-transition. For this configuration, one obtains the following system of equations:
\begin{equation}
\begin{aligned}
&\dot{\rho}_{\text{ss}} = -\frac{i}{2}  \Omega _{397} \left(\tilde{\rho} _{\text{sp}}-\tilde{\rho} _{\text{ps}}\right) + \Gamma_{\text{ps}} \rho_{\text{pp}}\\
&\dot{\rho}_{\text{dd}} = -\frac{i}{2}  \Omega _{866} \left(\tilde{\rho} _{\text{dp}}-\tilde{\rho} _{\text{pd}}\right) + \Gamma_{\text{pd}} \rho_{\text{pp}}\\
&\rho_{\text{ss}} + \rho_{\text{pp}} + \rho_{\text{dd}} = 1\\
&\dot{\tilde{\rho}}_{\text{sp}} = -i \Delta_{397} \tilde{\rho} _{\text{sp}} - \frac{1}{2} (\Gamma_{\text{ps}} + \Gamma_{\text{pd}}) \tilde{\rho} _{\text{sp}} - \frac{i}{2} \Omega_{866} \tilde{\rho} _{\text{sd}} + \frac{i}{2} \Omega_{397} (\rho_{\text{pp}}-\rho_{\text{ss}})\\
&\dot{\tilde{\rho}}_{\text{dp}} = -i \Delta_{866} \tilde{\rho} _{\text{dp}} - \frac{1}{2} (\Gamma_{\text{ps}} + \Gamma_{\text{pd}}) \tilde{\rho} _{\text{dp}} - \frac{i}{2} \Omega_{397} \tilde{\rho} _{\text{ds}} + \frac{i}{2} \Omega_{866} (\rho_{\text{pp}}-\rho_{\text{dd}})\\
&\dot{\tilde{\rho}}_{\text{sd}} = -i(\Delta_{397}-\Delta_{866}) \tilde{\rho} _{\text{sd}} + \frac{i}{2} \Omega_{397} \tilde{\rho} _{\text{pd}} - \frac{i}{2} \Omega_{866} \tilde{\rho} _{\text{sp}}\\
&\tilde{\rho}_{\text{ij}} = \tilde{\rho}_{\text{ji}}^{*}
\end{aligned}
\end{equation}
where $\tilde{\rho}_{\text{sp}} = e^{-i \omega_{397} t}\rho_{\text{sp}}$, $\tilde{\rho}_{\text{pd}} = e^{-i \omega_{866} t}\rho_{\text{pd}}$ and $\tilde{\rho}_{\text{sd}} = e^{-i (\omega_{397}-\omega_{866}) t}\rho_{\text{sd}}$.

\subsection{Intensity and visibility in the far field}
The far field intensity of the light emitted by two coherently driven atoms is given by \cite{agarwal}
\begin{equation}
I(\vec{r},t) \propto \sum_{i,j} \langle S_{+}^{(i)}(t)S_{-}^{(j)}(t)\rangle e^{-i(\vec{R}_i-\vec{R}_j)\cdot(\vec{k}_L-k_L\vec{n})}
\end{equation}
where $S_{+}^{(i)}=\ket{p_i}\bra{s_i}$, $S_{-}^{(i)}=\ket{s_i}\bra{p_i}$ and $\vec{R}_i$ denote the raising and lowering operators and position of atom $i$, $\vec{k}_L$ the wave vector of the driving laser and $\vec{n}$ the direction of observation. 
\\ 
In the steady state, one obtains:
\begin{equation}
\begin{aligned}
&\langle S_{+}^{(1)}S_{-}^{(1)}\rangle = \langle S_{+}^{(2)}S_{-}^{(2)}\rangle = \rho_{\text{pp}}\\
&\langle S_{+}^{(1)}S_{-}^{(2)}\rangle = \langle S_{-}^{(1)}S_{+}^{(2)}\rangle = \langle S_{+}^{(1)}\rangle \langle S_{-}^{(2)}\rangle = |\rho_{\text{ps}}|^2\\
&I(\vec{r}) \propto \left(2\rho_{\text{pp}}+|\rho_{\text{ps}}|^2 \left(e^{-i(\vec{R}_1-\vec{R}_2)\cdot(\vec{k}_L-k_L\vec{n})} + e^{i(\vec{R}_1-\vec{R}_2)\cdot(\vec{k}_L-k_L\vec{n}})\right)\right)\\
&\propto 2 (\rho_{\text{pp}}+|\rho_{\text{ps}}|^2 \cos{\phi})
\end{aligned}
\end{equation}
with
$\phi = (\vec{R}_1-\vec{R}_2)(\vec{k}_L-k_L\vec{n})$.
The visibility of the interference pattern is thus given by 
\begin{equation}
\mathcal{V} = \frac{|\rho_{\text{ps}}|^2}{\rho_{\text{pp}}}
\end{equation}
The steady state density matrix elements are obtained by setting the time derivatives to zero and solving the resulting linear system of equations. For the two level model, this yields

\begin{equation}
\begin{aligned}
&\tilde{\rho}_{\text{ps}} = \frac{i \Omega_{397}}{2(\frac{\Gamma_{\text{ps}}}{2}-i \Delta_{397})(1+S)}\\
&\rho_{\text{pp}} = \frac{S}{2(1+S)}
\end{aligned}
\end{equation}
where $S$ is the saturation parameter
\begin{equation}
S = \frac{\Omega_{397}^2/2}{\Delta_{397}^2 + \Gamma_{\text{ps}}^2/4}
\end{equation}
This results in a visibility
\begin{equation}
\mathcal{V} = \frac{1}{1+S} = \frac{1+4\Delta_{397}^2/\Gamma_{\text{ps}}^2}{1+4\Delta_{397}^2/\Gamma_{\text{ps}}^2+2 \Omega_{397}^2/\Gamma_{\text{ps}}^2}
\end{equation} 

In the three level model, the expression for the visibility reads 

\begin{equation}
\begin{aligned}
\mathcal{V} &= (\Omega_{866}^2 (\Gamma_{\text{pd}}^2 (4 \Gamma_{\text{ps}}^2 (\Delta _{397}-\Delta _{866})^2 + \Omega_{397}^4) +\\ 
&2 \Gamma_{\text{pd}} \Gamma_{\text{ps}} (4 \Gamma_{\text{ps}}^2 (\Delta _{397}-\Delta _{866})^2 + \Omega _{397}^2 (4 \Delta _{397} (-\Delta _{397}+\Delta _{866}) + \Omega_{866}^2 )) +\\
&\Gamma_{\text{ps}}^2 (4 \Gamma_{\text{ps}}^2 (\Delta _{397}-\Delta _{866})^2 + (4 \Delta _{397} (-\Delta _{397}+\Delta _{866}) + \Omega_{866}^2 )^2 ))) /\\
&((\Gamma_{\text{pd}} + \Gamma_{\text{ps}}) (\Gamma_{\text{pd}} \Omega_{397}^2 (4 (\Delta _{397}-\Delta _{866})^2 ((\Gamma_{\text{pd}} + \Gamma_{\text{ps}})^2 + 4 \Delta _{866}^2) + 8 (\Delta _{397}-\Delta _{866}) \Delta _{866} \Omega_{397}^2 + \Omega_{397}^4) +\\
&(4 \Gamma_{\text{ps}} ((\Gamma_{\text{pd}} + \Gamma_{\text{ps}})^2 + 4 \Delta _{397}^2) (\Delta _{397}-\Delta _{866})^2 + 8 (\Gamma_{\text{pd}} + \Gamma_{\text{ps}}) (\Delta _{397}-\Delta _{866})^2 \Omega_{397}^2 +\\
&(2 \Gamma_{\text{pd}} + \Gamma_{\text{ps}}) \Omega_{397}^4)  \Omega_{866}^2 + (8 \Gamma_{\text{ps}} \Delta _{397} (-\Delta _{397}+\Delta _{866}) + (\Gamma_{\text{pd}} + 2 \Gamma_{\text{ps}}) \Omega_{397}^2) \Omega_{866}^4 + \Gamma_{\text{ps}} \Omega_{866}^6 ))
\end{aligned}
\end{equation}

In Figure 1, the visibility in the three level model is shown for two different sets of laser parameters and compared to the visibility in the two level model. It can be seen that the additional decay channel towards the $\ket{d}$-state can both increase and decrease the visibility, depending on the laser parameters. 

\begin{figure}[h]
\includegraphics[width=0.6 \textwidth]{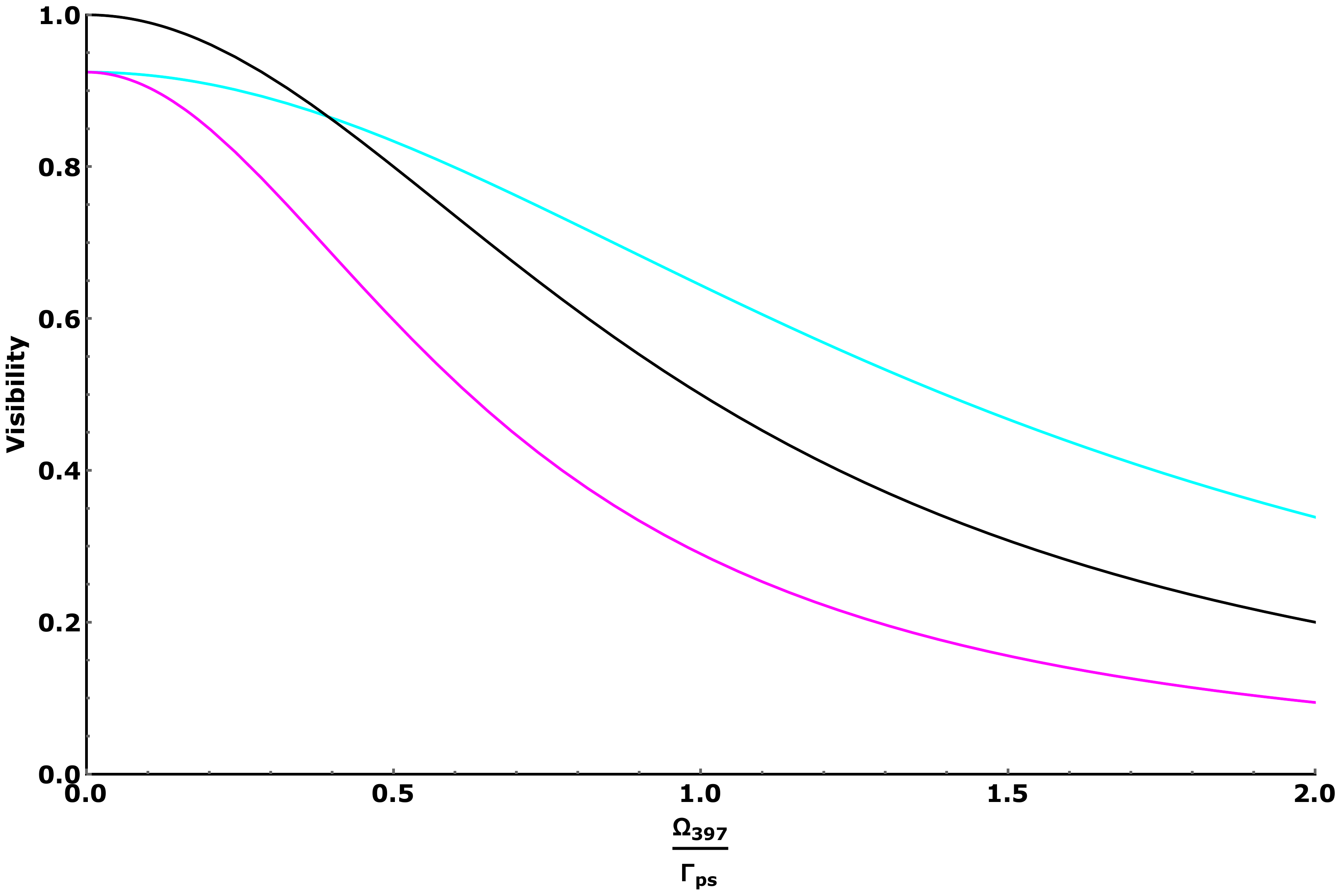}
\caption{The black curve shows the visibility in the two level model for a detuning of $\Delta_{397} = -10\,$MHz as a function of the Rabi frequency $\Omega_{397}$. The visibility as a function of   $\Omega_{397}$ in the three level system for two different Rabi frequencies of the repumping laser is shown by the cyan curve ($\Omega_{866} = 60\, \Gamma _\text{pd}$, $\Delta_{866} = +60\,$MHz) and the magenta curve ($\Omega_{866} = 20 \, \Gamma _{\text{pd}}$, $\Delta_{866} = +60\,$MHz). }
\end{figure}

\end{document}